\documentclass[manuscript]{aastex}

\shorttitle{Size Distribution of the Kuiper Belt's Scattering Objects}
\shortauthors{Shankman et al.}

\begin{document}

\title{A Possible Divot in the Size Distribution of the Kuiper Belt's Scattering Objects}

\author{C. Shankman, B. J. Gladman, }
\affil{Department of Physics and Astronomy, University of British Columbia, 
6224 Agricultural Road, Vancouver, BC, V6T 1Z1}

\author{N. Kaib}
\affil{Department of Physics and Astronomy, Queens University, Canada}

\author{J.J. Kavelaars}
\affil{Herzberg Institute of Astrophysics, National Research Council of Canada, Victoria, BC, Canada}

\and

\author{J.M. Petit}
\affil{Institut UTINAM, CNRS-Universit\'{e} de Franche-Comt\'{e}, Besan\c{c}on, France}

\begin{abstract}
Via joint analysis of a calibrated telescopic survey, which
found scattering Kuiper Belt objects, and models of their expected orbital
distribution, 
we explore the scattering-object size distribution.
Although for $D>$100~km the number of objects quickly rise as diameters
decrease, we find a relative lack of smaller objects, ruling out a single power-law at greater than 99\% confidence.
After studying traditional ``knees'' in the size distribution,
we explore other formulations and find that, surprisingly, our 
analysis is consistent with a very sudden decrease (a divot) in the number
distribution as diameters decrease below 100~km, which then rises again 
as a power-law. 
Motivated by other dynamically hot populations 
and the Centaurs, we argue for a divot size distribution where
the number of smaller objects rises again as expected 
via collisional equilibrium. 
Extrapolation yields
enough kilometer-scale scattering objects to supply the nearby Jupiter-Family comets. 
Our interpretation is that this divot feature is a preserved relic of
the size distribution made by planetesimal formation, now ``frozen in''
to portions of the Kuiper Belt sharing a ``hot'' orbital inclination
distribution, explaining several puzzles in Kuiper Belt science.
Additionally, we show that to match today's scattering-object inclination distribution, 
the supply source that was scattered outward must have already been vertically heated to of order 10$^{\circ}$.

\end{abstract}


\section{Introduction}
Measurements of the Kuiper Belt's size distribution (number at each diameter $D$) constrain accretional processes at planet formation and, potentially, subsequent collisional or physical evolution. 
Because astronomers observe brightnesses rather than $D$, object absolute magnitudes $H$ are tabulated as the observable proxy for the size distribution.
Collisional and accretional theories suggest exponential forms for the $N(H)$ distribution.
A differential number distribution of the form 
$\mathrm{d}N/\mathrm{d}H \propto 10^{\alpha H_g}$ 
with a logarithmic `slope' $\alpha$ corresponds to a power-law $D$ 
distribution $\mathrm{d}N/\mathrm{d}D \propto D^{-(5\alpha + 1)}$.
Although power-law $D$ distributions provide acceptable fits to Kuiper Belt surveys 
over spans of a few magnitudes in $H$, 
departures from single power-laws are necessary over larger $H$ ranges 
\citep{Jew98,Gla01,Ber04,Fue08,Fra08}.
For the steep ($\alpha$=0.8--1.2) distributions seen in the Kuiper Belt,
detections are dominated by objects near the largest $H$ magnitude (smallest size) visible in a given survey. 
An $\alpha > 0.6$ slope cannot continue as $H\rightarrow\infty$ ($D\rightarrow 0$~km) 
or the total mass diverges; thus a slope change (generically called a break) is 
required.
Evidence of such a break now exists for trans-Neptunian objects (TNOs) in 
the main Kuiper Belt (at distances $d \simeq$ 38--46 AU), both near the 
sensitivity limit of ground-based telescopic surveys 
\citep{Fra08,Fue08}
(reaching $H_g\sim$9--10 at 40~AU) and from deeper HST 
\citep{Ber04}
observations ($H_g\sim$13 at 40~AU).
This break has been modelled as a gradual transition to a smaller 
value of $\alpha$, a ``knee''. 

Probing 
a break is difficult because small TNOs are faint.
This problem is reduced when observing the scattering objects (SOs); 
these are mostly TNOs with perihelia $q$ $\le$ 35~AU 
(see below) and thus smaller objects are detectable 
while near the Sun.

At any time some SOs are only $d$ = 20--30 AU away, allowing a 4-m telescope, in excellent conditions, 
to detect objects down to $H_g \sim 12$.
For a monotonically increasing number distribution $N(H_g)$, the 
abundant small objects at the observable volume's innermost edge 
should dominate the detected sample. 
This is not what our survey found (Fig.~1), necessitating a relative lack of small SOs. 

We discarded a sudden (ad-hoc) albedo change, as it would produce a 
{\it gap} in $H$-space, 
not a drop, which does not match the observations; the
needed change is a sudden lack of $H_g > 9$ (and therefore small) SOs.

\section{Models}

Several models of the SO orbital element distribution
were exposed 
to the calibrated 
observational biases of the Canada France Ecliptic Plane Survey (CFEPS)
in order to quantitatively constrain the intrinsic $N(H_g)$ distribution. 
Drawing SOs from an orbital distribution model, and selecting $H_g$ from 
a candidate $N(H_g)$ distribution, the CFEPS survey simulator 
\citep{Jon06}
determines each object's observability and produces a set of 
``simulated detections'' expected from the model.

Two different SO orbital models are from a modified version of 
\citet{Kai11b}
(henceforth KRQ11).  
KRQ11 focuses on the effects of solar migration in the Milky Way on 
Oort Cloud structure.  
While this is not the focus of the current work, we can use the KRQ11 
control calculations, which assume an unchanging local galactic environment.

To test the sensitivity of our results to the 
dynamical context, we performed the same analysis 
on an independent model.
\citet{Gla06}
modelled the scattering of objects in an initial Solar System 
having an additional planet of order Earth mass.
As previously reported
\citep{Pet11}
this model also (perhaps surprisingly) 
satisfactorily
represents the current SO ($a,q$) distribution,
although too ``cold'' in inclinations.
In fact, this model and the cold KRQ11 model produced very similar results,
showing that our conclusions 
are mostly insensitive to the assumed Solar System history. 
Objects 
currently in the Centaur and detectable SO region (mostly $a<$200~AU) 
have, unsurprisingly, almost forgotten their initial state except 
for the the inclination 
distribution;
the current SO orbital distribution is not diagnostic of the
number and position of the planets early in the Solar System's
history.

\section{Observations}

CFEPS provided a set of
detections of outer Solar System objects in a precisely calibrated
survey
\citep{Jon06,Kav09}
whose pointing history,
detection efficiency, and  tracking performance were recorded.
The final set of TNO detections (with full high-precision orbits)
and the fully calibrated pointing history make up the L7 release
\citep{Pet11}.
This absolute calibration of CFEPS allows a model of the present 
orbital (and size) distribution of to be passed through the CFEPS Survey Simulator,
yielding a set of simulated detections whose orbital and $H_g$
distributions can be compared to the real detections.

The three
models provide orbital distributions of all TNOs.
The ``scattering'' TNOs are then selected out of the final 10 Myr  
stage of the model integrations using the 
criteria: variation of $a > $1.5~AU in semimajor axis during 10 Myr,
with $a<1000$~AU
\citep{Mor04,Gla08}.
Historically, a simple $q$ cut was used to isolate the ``scattered disk''
\citep{Dun97,Luu97,Tru00}, which 
has serious
disadvantages when trying to discuss the cosmogony, as there is
a nearly impossible distinction between implanted (and thus
scattered) and the original Kuiper Belt population (if any).
Perihelion divisions also 
undesirably includes resonant objects and most inner 
main-belt TNOs.

The CFEPS SO sample consists of 9 objects 
(Table 1), supplemented by two SOs 
discovered in a high-latitude extension survey 
(covering $\simeq 470$ sqdg in 2007--2008, extending up to 
65$^\circ$ ecliptic latitude), which was fully calibrated in 
the same way as CFEPS.

To characterize the form of the $N(H_g)$, we introduce a novel 
formulation, allowing 
for the exploration of distributions with knees and divots (a sudden 
drop in the differential number of objects followed by a recovery). 
We parameterised the 
$H_g$
distribution (Fig.~2A) with the fixed slope $\alpha_b = 0.8$ (see below) 
for SOs brighter than a break at $H_g = 9$ 
($D \simeq$100~km), allowed an adjustable slope $\alpha_f$ for fainter 
objects, and an adjustable contrast $c \ge 1$. 

The $H_g$-magnitudes are drawn from one of three types of distributions: 

\noindent
{\bf(1)} a single exponential of logarithmic slope $\alpha$, 
{\bf(2)} $N(H_g$) with a knee (contrast $c$ =1). That is, one 
slope $\alpha_{b}$ for SOs with $H_g<H_{knee}$ 
and $\alpha_{f}$ for $H_g>H_{knee}$, 
where $N(H_g)$ 
is continuous across the knee at $H_{knee}$ and negative slopes $\alpha_{f}$ 
are allowed as 
suggested 
\citep{Ber04}, and 
{\bf(3)} one slope $\alpha_{b}$ to a divot at $H_{divot}$,
which is a sudden drop in differential number by a factor $c$, 
with a potentially different slope $\alpha_{f}$ beyond the cliff
at $H=H_{divot}$. 
Although in reality the discontinuity is unlikely to be an 
instantaneous drop, our data do not merit trying to constrain 
the values of the expected steep negative slope and small extent over which it 
drops;  collisional models 
\citep{Fra09,Cam12}
do show collisional divots where the drop occurs over 
$D$ ranges of factors $<2$ 
(a few tenths of magnitude in $H_g$).

In principle there are four parameters: $\alpha_{b}$, $\alpha_{f}$, $H_{divot}$, and $c$ (Fig. 2).
For $H_g<9$ a single power-law of $\alpha_{b}\simeq0.8$ does 
indeed match our detections; we elected 
to fix this slope at that value with the unifying philosophy that all 
the hot transneptunian populations share this same hot slope;
$\alpha_{b}=0.8$  matches both
the hot Classical belt measured down to $H_g\simeq8.0$ 
\citep{Pet11,Fra08,Fue08},
and to the 3:2 resonators measured down to $H_g\simeq9.0$
\citep{Gla12}.
Our detections require a transition around $H_g= $9--10
to explain the relative lack of small 
detections;
we thus fixed the knee/divot for our analysis at $H_g = 9$ 
(slightly larger than $D$=100 km for 5\% $g$-band albedo).
This leaves only two free parameters: the contrast $c$ at the divot 
and the slope $\alpha_f$ for 
absolute magnitudes fainter than the divot/knee.

To assess a match, the Anderson-Darling (AD) statistic is calculated between 
our 11-object sample and the distribution of simulated detections from 
the model, for each orbital parameter. 
An AD significance level of $<5\%$ 
rejects
the hypothesis that the real SO observations could be drawn from 
the simulated detections at the $95\%$ confidence level (for that 
orbital parameter). 
To retain a model, we required
that none of the $q$, $d$, $i$, and $H_g$ distributions 
are
rejectable at $> 95\%$ confidence.

\section{Absolute Magnitude Distribution}

The observational bias is strong 
(Fig.~1) 
, but when accurately calibrated allows us to constrain the 
$H_g$ distribution's form.
Single power-laws predict significantly more close-in detections 
than were seen by CFEPS; for a slope of $\alpha = 0.8$, roughly half 
of the expected detections (Fig.~1 D's blue dashed curve) should have 
a distance at detection $d <$ 23 AU, which is the closest real 
SO in our sample.  
The observationally biased models predict 
that the majority of detected SOs would have orbits with $q < $20~AU  
at $d$ = 20--25 AU and be small ($H_g > 9$ or $D \le 100$ km) objects, 
in contrast to our detections, which demonstrates that our observations 
are sensitive beyond the break. 
When confined to $q>25 AU$ (where objects
must be large to be seen) the orbital models provide good matches, however
extensions to smaller distance fail
when using a single power-law, 
pointing to a breakdown arising from the the assumed $N(H_g)$. 

We rule out a single power-law of slope 0.8 at 99\% confidence, and 
can rule out all single power-laws with slope between 0 and 1.2 
at 95\% confidence. 
Slopes of 0.5 and 0.6 are not rejectable across the whole distribution, but 
are rejectable (95\% confidence) when the distribution is considered 
in both $H_g > 9$ and $H_g < 9$ subsets; 
we demand these work because 
the steep slopes measured for other hot populations match our 
$H_g < 9$ detections well, and a 
shallower
slope is 
erroneously
found by measuring across a divot feature when requiring a single slope 
(see below).

Our relatively small sample is powerful because our detected SOs span the 
break and, 
when coupled with the precise CFEPS calibration, allows the
non-detection of $H_g=$10-12 SOs (several magnitudes past the 
divot) to provide a strong constraint on $N(H_g)$.
Down to this limit,
CFEPS detected moving objects as close as 20 AU with no rate of motion 
dependence.
Because our orbits are accurate, we can separate the SOs
from the other hot populations, and use a dynamical model specific
to the SOs. 

All of our $N(H_g)$ cases have the obvious and previously-known problem that the
model's orbital inclinations are mostly lower than the true population's
\citep{Pet11,Gla12}, even for the cases where the $N(H_g)$ otherwise provides a good match.
For example, Fig.~1 shows a divot ($c \simeq 6$, $\alpha_f$ = 0.5) 
producing 
a good match between the model's expected detections (green curve) and 
the real SO sample (red), excepting the $i$ problem. 
Models 
\citep{Lev08}
which scatter out a cold TNO population (from $d<30$~AU 
with initial inclination distribution widths $\sigma_i \leq 6^{\circ}$) 
to eventually form today's hot population produce 
current TNO populations where too many low-$i$ detections are expected 
in observational surveys (Fig.~1B). 
This is part of growing evidence that the original planetesimal disk supplying 
today's high-$i$ objects must have already been vertically excited
before being scattered out 
\citep{Pet11,Gla12,Bra12},
which has strong cosmogonic implications for an extended 
quiescent phase of the early Solar System
\citep{Lev08}.
We therefore computed a new SO model with a hotter 
($\sigma_i \simeq 12^{\circ}$) initial disk; 
this provides an excellent match with today's SO $i$-distribution
(see Fig.~3)
and we constrain $N(H_g)$ below using this model.

To constrain the size distribution, a grid of possible divot contrasts and post-divot slopes was explored. 
Fig.~4 shows acceptability levels for the range of explored parameter pairs
($c, \alpha_f$).
A single power-law of $\alpha = 0.8$ (blue star Fig. 4) 
has $<1$\% probability. 
We are left with a range of acceptable parameter space, including knee ($c$=1) and divot ($c >$1) scenarios; we further constrain $N(H_g)$ by looking to other Kuiper Belt populations.

The so-called hot Kuiper Belt populations (the hot main belt, inner belt, resonant, and detached TNOs) share an $i$ distribution half-width of roughly $15^\circ$ 
\citep{Pet11}
with the SOs, suggesting a cosmogonic link. 
In analyses of deep luminosity functions dominated by hot main-belt detections,
the common conclusion 
\citep{Ber04,Fue10}
was that for magnitude $g>25$ the slope must break to a 
faint $\alpha_f<0.3$ value or even become negative in order to explain the
lack of detections in the following few magnitudes; beyond this 
no data exists for the main Kuiper Belt.
For SOs and their companion objects (Centaurs), however, many $H_g\gg9$
objects are known from wide-field surveys, mostly detected at 
$d<20$ AU.
In fact, measurements of Jupiter Family Comets (JFCs) in the
$H_g\approx$14--17 range give slopes $\alpha_f\simeq0.5 \pm 0.1$
(see Table~6 of \citet{Sol12}).
These two arguments mean 
the SO distribution cannot remain at $\alpha_f<0.3$.
leading us to discard knees to negative slopes.
A divot can explain both a relative lack of objects beyond the break and the eventual recovery necessary to provide the JFCs. 
A divot also motivates the negative slopes measured, as a realistic divot will take the form of a decrease at the break, rather than the sharp discontinuity we use. 
We prefer the divot solution with $\alpha_f=$0.5 and $c\simeq$6 (green star Fig.~4) which 
matches the observations and allows for a single slope $\alpha_f=0.5$ 
from the divot out to the $H_g>$14 Jupiter Family comets whose slope 
is near the collisional equilibrium value 
\citep{Obr05}.

\section{External Arguments}

As the hot populations have similar colours and $D>100$-km size distributions, it seems likely that
they were all transplanted to join a pre-existing 
cold Kuiper Belt 
\citep{Pet11}
during a common event early in the Solar System's history, and would 
thus logically share the same divot and small $D$ distribution.
Such a transplant process 
can
successfully implant TNOs in the stable Kuiper Belt 
\citep{Lev08,Bat11},
although the resonant population ratios and $i$ distribution
are problematic
\citep{Gla12}.
We thus look for evidence of such a feature in other hot populations.

\subsection{The Neptune Trojans}

A search for Neptune Trojans 
\citep{She10}
provided significant evidence that an $\alpha\sim0.8$ power-law cannot
continue for $D <$ 100-km Trojans; a divot was not apparent because 
Trojans significantly smaller were not detected. 
The dispersed 
Trojan inclination distribution 
\citep{She06}, 
although not yet precisely measured, 
links these objects to all the other resonant populations 
\citep{Gla12}. 
\citet{She10} 
used Neptune Trojan searches to
argue that beyond $m_R\simeq$23 (corresponding to $H_g\simeq9$) there
was an absence of Trojans due to non-detections,
and thus smaller Trojans were missing.
Assuming that the Trojans and other resonant TNOs were
implanted from a scattering population, and thus share the
same size distribution, we confirmed that our divot $N(H_g)$
matches the lack of $D<$100 km Neptune Torjan detections in the \citet{She10} surveys.

Given our analysis, the conclusion would not be that small Neptune
Trojans are ``missing'', but rather that the sudden drop 
results in the population fainter than the divot not
recovering in on-sky surface density until at least $H_g>11$,
by which point the deepest survey lacked the sensitivity
to detect them.
If correct, detection of several small ($H_g>11$)
Trojans requires surveying $\sim100$ square degrees 
to 26th magnitude at the correct elongation.

\subsection{Hot Populations}

A recent deep telescopic survey 
\citep{Fra10}
estimated $\alpha \simeq 0.40 \pm 0.15$ (within error of our preferred $\alpha_f$ = 0.5) 
from the apparent-magnitude distribution for ``close'' ($30< d <38$) TNOs (orbits were not obtained). 
These distances are dominated by several hot populations, 
but the measurement is shallower than the usual hot population slope of 0.8. 
We calculated $H_g$ magnitudes for the \citet{Fra10} 
detections
and find that due to the survey's 
depth, this sample is dominated by $H_g>9$ TNOs and 
thus 
would
measure the post-divot slope.

\section{Feasibility of a Divot}

A primordial size-distribution wave at small sizes ($D$ = 2 km) could
propagate 
\citep{Fra09}
to $D\sim$100~km 
in a dynamically hot ($\Delta v = 2$ km/s) collisional 
environment after 500 Myr.
Alternately, recent modeling of planetesimal creation 
\citep{Joh07,Mor09},
suggests that the protosolar nebula may only have produced planetesimals 
larger than a certain critical diameter,
in which case the $\alpha_b=0.8$ slope and the $D\sim100$~km divot size 
are set by planetesimal formation physics; smaller objects appear only later 
due to collisional fragmentation. 
These scenarios match our results, where one interprets the hot population's 
$N(H_g)$ to have been ``frozen'' when suddenly transplanted (scattered) from 
a denser region nearer the Sun to the large volume it now 
occupies, ending the collisional evolution. 
An exciting prospect is that the divot directly records a preferential 
$D$ that planet building produced in the solar-nebula region where 
the hot TNOs originally formed, and that the divot's depth (which could
easily range from $c$=2--30) measures the integrated collisional evolution
(depending on both the duration of the pre-scattering phase and the 
random speeds present).
An initial distribution with no $D<$100~km TNOs was shown 
\citep{Cam12}
to evolve into a divot with $c \sim$ 20 and $\alpha_f \simeq 0.5$,
in the dynamical environment of the Nice model;
such a 500-Myr quiescent phase 
\citep{Gom05}
allows a divot to form but the evidence we find for a higher-$i$ 
early phase may argue instead for a much shorter and more intense collisional 
environment.

Our divoted $N(H_g)$ produces a cumulative distribution (Fig.~2 B)
with a shallow plateau for $H_g$=9--12, similar to that 
deduced for the hot population and SOs in deep HST
observations
\citep{Ber04,Vol08}
and estimated for scattering impactors of the saturnian moons
\citep{Min12}.
Our estimate of $2\times10^6$ SOs with $H_g<13$ and a slope of $\alpha_f=0.5$ 
extrapolates 
to $\sim2\times10^9$ SOs with 
$H_g<$18, providing a sufficient number 
\citep{Dun97,Vol08}
of SOs to feed the Jupiter Family Comets, while satisfying the observed plateau.

Single power-laws that fit the $H_g<9$ SOs fail when extended to 
smaller objects.
Our novel divot parameterisation (Fig. 2) matches our data and 
would
simultaneously explain the puzzles of the JFC source, 
the ``missing'' Neptune Trojans, 
and the known rollover in the Kuiper Belt's luminosity function.
To better constrain the form of the 
break,
a new survey 
must find and determine orbits for $\sim$10 SOs from 10--30 AU;
this requires discovery (and tracking over several degrees of arc) 
targets moving up to 15''/hr by observing $\sim$200 sq.~deg. to 
24th magnitude.

{\it Facilities:} \facility{CFHT}, \facility{NRC (HIA)}

\appendix

\clearpage

\begin{figure}
\includegraphics[width=1.01\textwidth]{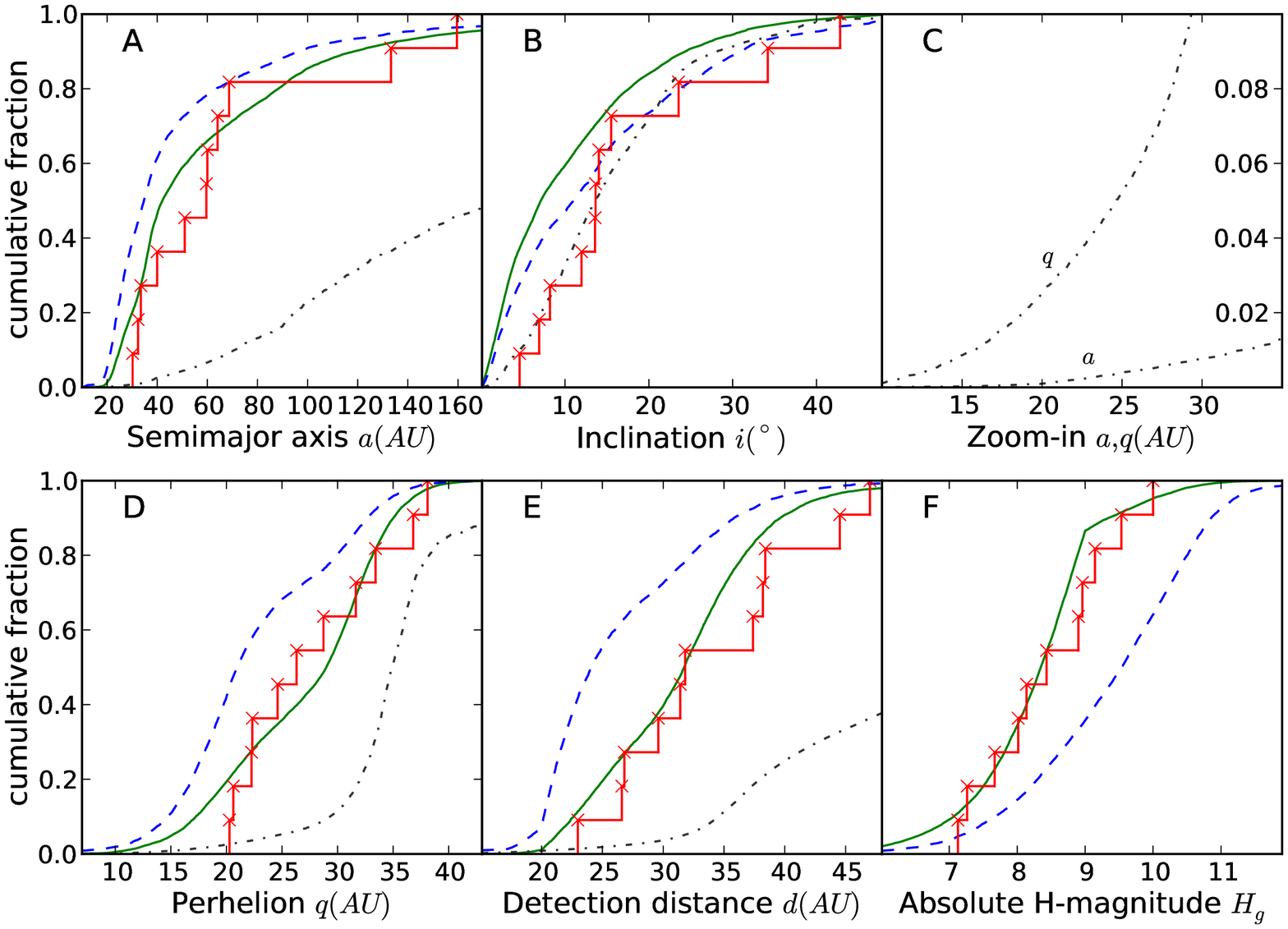}\label{cumulatives}
\caption{Cumulative distributions for 5 quantities, comparing the 
observed objects (red staircase) with the initially cold model's simulated detections.
The black dash-dot curves show the orbital model's intrinsic 
$a, i$, $q$, and $d$ distributions.
When coupled to two different $N(H_g)$ distributions, the biases produce the 
differing predictions for the detections. 
A single power-law slope of $\alpha = 0.8$ (blue dashed) is 
rejectable at $>$ 99\% confidence 
in $d$, $q$ and $H_g$.  
Contrastingly, our preferred divot $N(H_g)$ 
(green curve, see Fig. 2) provides vastly better matches, although both produce too 
many low-$i$ detections.
}
\end{figure}
\clearpage
\begin{figure}
\includegraphics[width=1\textwidth]{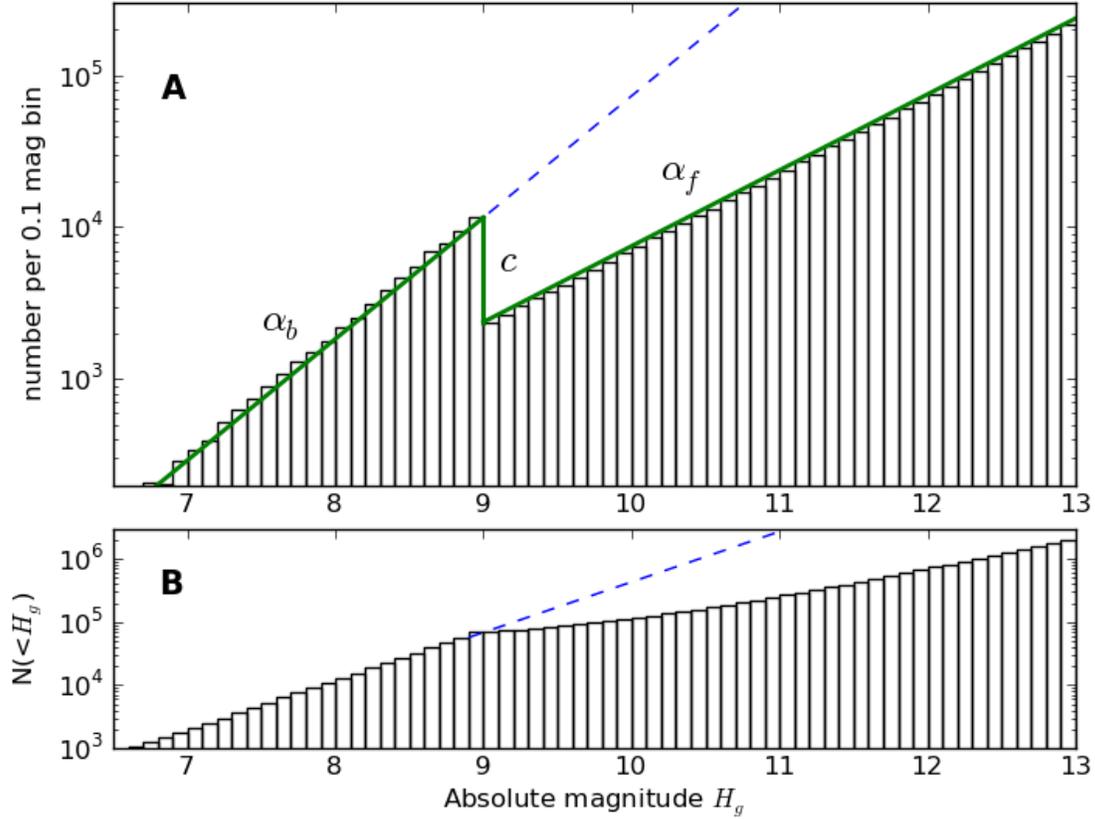}\label{hdits}
\caption{
Histograms of the $N(H_g)$ distribution for our preferred 
divot solution. 
The vertical axes show the total SO numbers using the absolute CFEPS calibration \citep{Pet11}.
{\bf A:} The differential distribution (solid green), with an extrapolated
$\alpha =$ 0.8 beyond $H_g$ = 9 (dashed blue).
The contrast $c \simeq 6$ is the ratio of differential number on either 
side of the divot.  
{\bf B:} The cumulative version.
For $H_g>$13 the cumulative $N(<H_g)$ has reached slope $\alpha_f$ after 
the flattened region following the divot. 
}

\end{figure}
\clearpage

\begin{figure}
\includegraphics[width=1\textwidth]{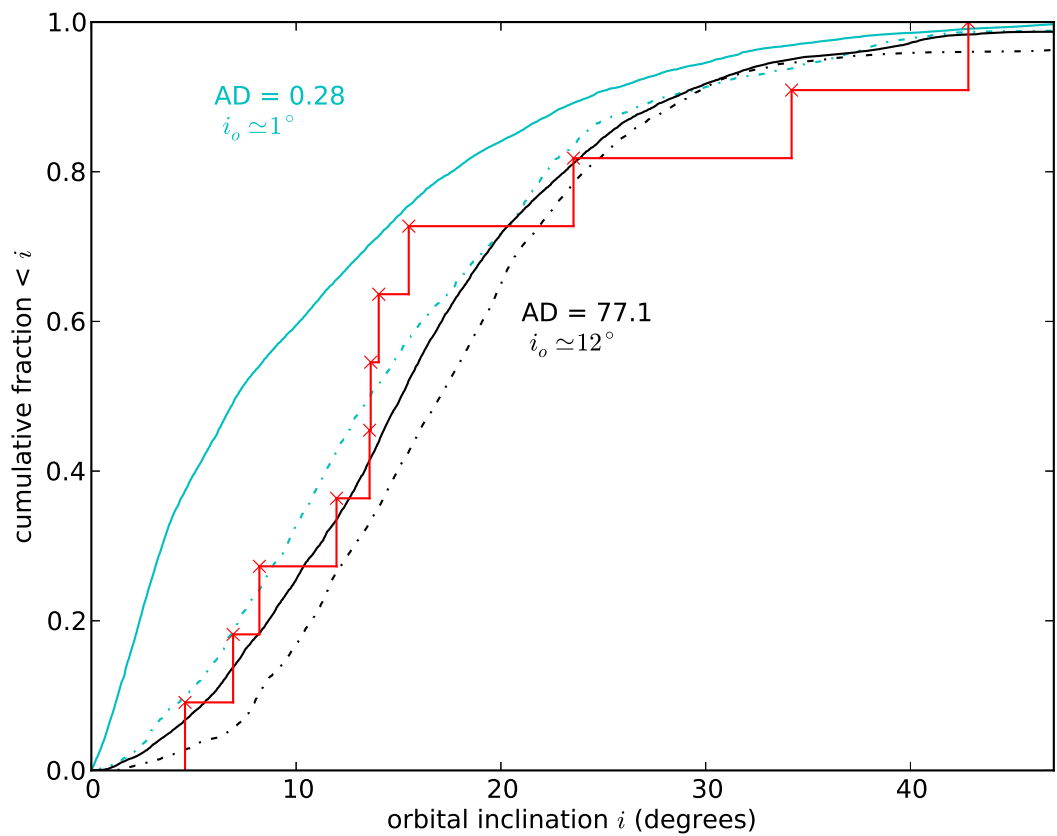}\label{incd}
\caption{The hot (black dashed) and cold (light blue dashed) 
KRQ11 intrinsic $i$ distributions produce the biased distributions (solid lines) from our preferred model
for comparison with the CFEPS sample (red). 
The hot model significantly improves the match because it 
has a relative lack (at the current epoch)  of low-$i$ SOs
to be detected by the survey; the eleven CFEPS SOs 
have $<1$\% probability of being drawn from the
initially cold simulation.
}
\end{figure}

\clearpage
\begin{figure}
\includegraphics[width=0.98\textwidth]{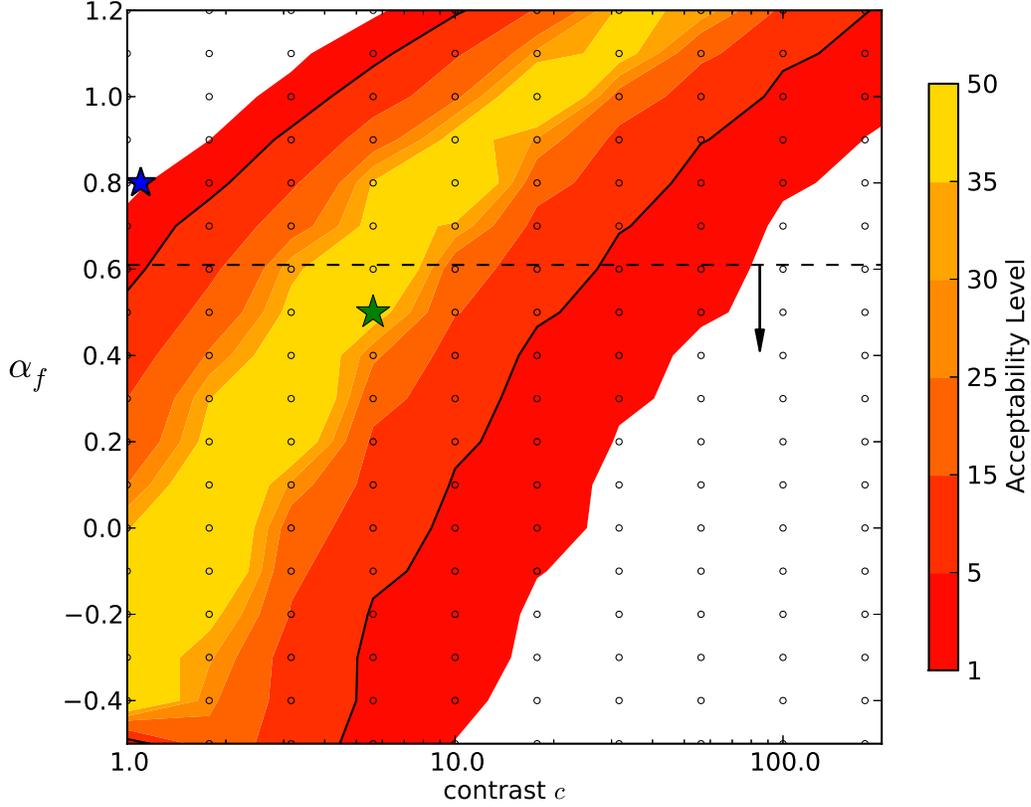}\label{contours}
\caption{Contour values (colours) computed from a grid (points) in ($c,\alpha_f$) space, giving the Anderson-Darling probability of drawing the 
most rejectable of the $d$, $q$, $i$, and $H_g$ distributions from 
the hot KRQ model of today's SOs.
White areas indicate when 
the ($c$,$\alpha_f$) pair had 
$<$1\% probability of 
coming from the model, and the black contour bounds $<$5\%.  
A single power-law (blue, darker star) is rejected.
Our favoured model (green, lighter star) satisfies both 
(a) $\alpha_f\simeq0.5$ like known JFCs, 
(b) $\alpha \leq 0.6$ which prevents the extrapolated mass of small 
SOs from diverging.
}
\end{figure}

\clearpage

\begin{table}
\begin{center}
\begin{tabular}{| c | c | c | c | c | c |}
\hline
\bf{Designation} & \bf{a (AU)} & \bf{q (AU)} & \bf{i (deg)} & \bf{d (AU)} & \bf{H$_g$} \\
\tableline                                                                                                                                                                                                                                      
L4k09 & 30.19 & 24.60 & 13.586 & 26.63 & 9.5 \\
HL8a1 & 32.38 & 22.33 & 42.827 & 44.52 & 7.3 \\
L4m01 & 33.48 & 28.73 & 8.205  & 31.36 & 8.9 \\
L4p07 & 39.95 & 26.31 & 23.545 & 29.59 & 7.7 \\
L3q01 & 50.99 & 33.41 &  6.922 & 38.17 & 8.1 \\
L7a03 & 59.61 & 22.26 &  4.575 & 46.99 & 7.1 \\
L4v11 & 60.04 & 31.64 & 11.972 & 26.76 & 10.0 \\
L4v04 & 64.10 & 38.10 & 13.642 & 31.85 & 9.1 \\
L4v15 & 68.68 & 36.81 & 14.033 & 22.95 & 9.0 \\
L3h08 & 159.6 & 20.26 & 15.499 & 38.45 & 8.0 \\
HL7j2 & 133.25& 20.67 & 34.195 & 37.38 & 8.4 \\

\tableline
\end{tabular}
\end{center}
\caption[CFEPS actively scattering sample.  Give MPC designations?]{\label{Table:CFEPSas} CFEPS + extension SO sample}
\end{table}


\begin{thebibliography}{}



\bibitem [Bernstein et al.(2004)]{Ber04} Bernstein, G.M. et al.  2004, \aj, 128, 1364 
\bibitem [Brasser \& Morbidelli (2012)]{Bra12} Brasser, R. \& Morbidelli, A.  2012 Asteroids, Comets, and Meteors meeting, abstract.
\bibitem [Batygin et al.(2011)]{Bat11} Batygin, K., Brown, M. \& Fraser, W.  2011, \apj, 13, 738
\bibitem [Campo Bagatin \& Benavidez(2012)]{Cam12} Campo Bagatin \& A.,Benavidez, P.  2012, \mnras, 423, 1254
\bibitem [Duncan \& Levison(1997)]{Dun97} Duncan, M. \& Levison, H.F.  1997, Science, 276, 1670
\bibitem [Elliot et al.(2005)]{Ell05} Elliot, J., Kern, S.D. \& Clancy, K.B.  2005, \aj, 129, 1117
\bibitem [Fraser(2009)]{Fra09} Fraser, W.  2009, \apj, 706, 119 
\bibitem [Fraser et al.(2010)]{Fra10} Fraser, W., Brown, M. \& Schwamb, M.  2010, Icarus, 210, 944
\bibitem [Fraser \& Kavelaars(2008)]{Fra08} Fraser, W. \& Kavelaars J.J.  2008, Icarus, 198, 452
\bibitem [Fraser \& Kavelaars(2009)]{FraK09} Fraser, W. \& Kavelaars, J.J.  2009, \aj, 137, 72
\bibitem [Fuentes \& Holman(2008)]{Fue08} Fuentes C., \& Holman, M.  2008, \aj, 136, 83
\bibitem [Fuentes et al.(2010)]{Fue10} Fuentes, C., Holman, M., Trilling, D. \& Protopapas, P.  2010, \apj, 722, 1290
\bibitem [Gladman et al.(2001)]{Gla01} Gladman, B. et al.  2001, \aj, 122, 1051 
\bibitem [Gladman \& Chan(2006)]{Gla06} Gladman, B. \& Chan, C.  2006, \apjl, 643, L135
\bibitem [Gladman et al.(2009)]{Gla09} Gladman, B., Kavelaars, J.J., Petit, J.-M., et al.  2009, \apjl, 697, L91
\bibitem [Gladman et al.(2012)]{Gla12} Gladman, B., Lawler, S., Petit, J.-M, et al. 2012, \aj, 144, 23
\bibitem [Gladman et al.(2008)]{Gla08} Gladman, B., Marsden, B.G. \& Vanlaerhoven, C.  2008, in The Solar System Beyond Neptune 43
\bibitem [Gomes et al.(2005)]{Gom05} Gomes, R., Levison,H.F., Tsiganis \& K., Morbidelli, A.  2005, Nature, 435, 466
\bibitem [Jewitt et al.(1988)]{Jew98} Jewitt, D., Luu, J., \& Trujillo, C.  1988,  \aj 115, 2125
\bibitem [Johansen et al.(2007)]{Joh07} Johansen, A., Oishi, J., Mac Low, M.-M., et al.  2007, Nature, 448, 1022
\bibitem [Jones et al.(2006)]{Jon06} Jones, R.L., Gladman, B., Petit, J.-M., et al.  2006, Icarus, 185, 508 
\bibitem [Kaib et al.(2011a)]{Kai11a} Kaib, N., Quinn, T. \& Brasser, R.  2011, \apj, 141, 3
\bibitem [Kaib et al.(2011b)]{Kai11b} Kaib, N.,  Ro\u{s}kar, R. \& Quinn, T.  2011, Icarus, 215, 491
\bibitem [Kavelaars et al.(2009)]{Kav09} Kavelaars, J.J., Jones, R.L, Gladman, B., et al.  2009, \aj, 137, 491
\bibitem [Levison et al.(2001)]{Lev01} Levison, H.F., Dones, L. \& Duncan, M.  2001, \aj, 121, 2253
\bibitem [Levison et al.(2008)]{Lev08} Levison, H., Morbidelli, A., Van Laerhoven, C., Gomes, R. \& Tsiganis, K.  2008, Icarus, 196, 258
\bibitem [Luu et al.(1997)]{Luu97} Luu, J., Marsden, B., Jewitt, D., et al.  1997, Nature, 387, 573
\bibitem [Minton et al.(2012)]{Min12} Minton, D., Richardson, J., Thomas, P., Kirchoff, M. \& Schwamb, M.  2012, in ACM Conf. Sess 551
\bibitem [Morbidelli et al.(2009)]{Mor09} Morbidelli, A., Bottke, W.F., Nesvorn\'{y}, D. \& Levison, H.  2009, Icarus, 204, 448
\bibitem [Morbidelli et al.(2004)]{Mor04} Morbidelli, A., Emel'yanenko, V. \& Levison, H.F.  2004, \mnras, 355, 935
\bibitem [O'Brien \& Greenberg(2005)]{Obr05} O'Brien, D. \& Greenberg, R.  2005, Icarus, 178, 179
\bibitem [Petit et al.(2011)]{Pet11} Petit, J.-M, Kavelaars, J.J., Gladman, B., et al.  2011, \aj, 142, 131
\bibitem [Sheppard \& Trujillo(2006)]{She06} Sheppard, S. \& Trujillo, C.  2006, Science, 313, 511
\bibitem [Sheppard \& Trujillo(2010)]{She10} Sheppard, S. \& Trujillo, C.  2010, \apjl, 723, L233
\bibitem [Solontoi et al.(2012)]{Sol12} Solontoi, M., Ivezi\'{c}, \v{Z}., Juri\'{c}, M. et al.  2012, Icarus, 218, 571
\bibitem [Volk \& Malhotra(2008)]{Vol08} Volk, K. \& Malhotra, R.  2008, \apj, 687, 714
\bibitem [Trujillo et al.(2000)]{Tru00} Trujillo, C., Jewitt, D., \& Luu, J.  2000, \apj, 529, 103





\end{thebibliography}
\end{document}